\documentclass[a4paper,fleqn]{cas-sc}
\usepackage{amsmath}
\usepackage{dsfont}
\usepackage[authoryear,longnamesfirst]{natbib}
\usepackage{threeparttable}
\usepackage{algorithmic}  
\usepackage{algorithm} 
\usepackage[algo2e]{algorithm2e}
\usepackage{graphicx}
\usepackage{float}
\def\tsc#1{\csdef{#1}{\textsc{\lowercase{#1}}\xspace}}
\tsc{WGM}
\tsc{QE}
\tsc{EP}
\tsc{PMS}
\tsc{BEC}
\tsc{DE}
\usepackage{dsfont}
\usepackage{bbding}
\usepackage{multirow}
\usepackage[normalem]{ulem}
\useunder{\uline}{\ul}{}

\begin{document}
\let\WriteBookmarks\relax
\def\floatpagepagefraction{1}
\def\textpagefraction{.001}

\shorttitle{Preprint}

\shortauthors{Yao et~al.}

\title [mode = title]{Adversarial Contrastive Domain-Generative Learning for Bacteria Raman Spectrum Joint Denoising and Cross-Domain Identification}                      


\affiliation[1]{organization={
State Key Laboratory of Precision Measurement Technology and Instruments, Tsinghua University}, 
    city={Beijing},
    postcode={100084}, 
    country={China}}

\author[1]{Haiming Yao}[style=chinese, orcid=0000-0003-1419-5489]
\ead{yhm22@mails.tsinghua.edu.cn}

\author[1]{Wei Luo}[style=chinese]
\ead{luow23@mails.tsinghua.edu.cn}

\author[1]{Xue Wang}[style=chinese]
\cormark[1]
\ead{imsnwangxue@gmail.com}

\cortext[cor1]{Corresponding author}

\begin{abstract}
Raman spectroscopy, as a label-free detection technology, has been widely utilized in the clinical diagnosis of pathogenic bacteria. However, Raman signals are naturally weak and sensitive to the condition of the acquisition process. The characteristic spectra of a bacteria can manifest varying signal-to-noise ratios and domain discrepancies under different acquisition conditions. Consequently, existing methods often face challenges when making identification for unobserved acquisition conditions, i.e., the testing acquisition conditions are unavailable during model training. In this article, a generic framework, namely, an adversarial contrastive domain-generative learning framework, is proposed for joint Raman spectroscopy denoising and cross-domain identification. The proposed method is composed of a domain generation module and a domain task module. Through adversarial learning between these two modules, it utilizes only a single available source domain spectral data to generate extended denoised domains that are semantically consistent with the source domain and extracts domain-invariant representations. Comprehensive case studies indicate that the proposed method can simultaneously conduct spectral denoising without necessitating noise-free ground-truth and can achieve improved diagnostic accuracy and robustness under cross-domain unseen spectral acquisition conditions. This suggests that the proposed method holds remarkable potential as a diagnostic tool in real clinical cases.

\end{abstract}


\maketitle

\section{Introduction}
\label{sec:introduction}

Bacterial infections result in a significant number of deaths and economic burdens. With the widespread use of antibiotics, drug-resistant strains continue to emerge, rendering the diagnosis and treatment of bacterial infections increasingly challenging\cite{r1,r2}. The accurate identification of bacterial species and their susceptibility to antibiotics is of widespread clinical interest, aiming to precisely diagnose the infecting bacterial species causing disease. Currently utilized diagnostic methods necessitate a slow sample culture process for the detection and identification of bacteria and their antibiotic sensitivities, a procedure that may span several days\cite{r3}. Developing novel methods for the rapid, accurate, and culture-free identification of bacterial infections is urgently needed to facilitate early and precise treatment.

Existing rapid microbial detection methods\cite{r4} have not been widely adopted in clinical practice due to their requirement for expensive equipment and the expertise of knowledgeable professionals. Raman microspectroscopy, as a form of label-free vibrational spectroscopy, when combined with a confocal optical configuration, allows for the examination of the biochemical characteristics of bacteria at the single-cell level\cite{r1,r5}. Bacteria of different categories possess distinct molecular compositions, and consequently, they exhibit different Raman spectra, serving as their biological phenotype. Thus, it demonstrates the potential to identify bacterial species and antibiotic resistance.

To achieve automated and efficient spectral analysis, researchers have employed various machine-learning techniques to handle various tasks in Raman spectroscopy\cite{r6}. Most of these techniques are utilized to eliminate redundant information from spectra and retain the most pertinent details. For example, due to inefficient Raman scattering, subtle spectral differences among bacteria are easily overshadowed by background noise, resulting in a low signal-to-noise ratio(SNR) of the acquired spectra. Researchers have employed various methods for noise removal\cite{r7}, with common approaches including principal component analysis (PCA)\cite{r8}, wavelet transform\cite{r9}, and artificial neural networks(ANN)\cite{r10}. Higher-level semantic tasks mainly involve spectrum classification tasks\cite{r11}. Due to subtle spectral differences among bacteria of different categories and variability within the same category, traditional discriminative machine learning methods such as linear discriminant analysis(LDA)\cite{r12}, support vector machine(SVM)\cite{r13}, and k-nearest neighbors(KNN)\cite{r14} perform moderately in complex spectral analysis tasks.

Recently, due to the advancements in deep learning, various deep neural networks(DNNs) have been applied to spectral analysis\cite{r37}. For instance, in \cite{r7}, DNN was employed to achieve high-throughput molecular imaging. In \cite{r15}, the generative adversarial network(GAN) was proposed and utilized for denoising Raman spectra. In \cite{r1}, a large-scale bacterial spectrum dataset was constructed, and a 1-dimensional convolutional neural network(CNN) was proposed to accurately identify 30 common bacterial pathogens. This method was further enhanced in \cite{r16} by incorporating a scale-adaptive network model. Additionally, in \cite{r17}, 2D Raman figures and CNNs were combined for Raman spectrum classification.

However, spontaneous Raman scattering is inherently a weak process, with a detection probability of only about $10^{-7}$\cite{r18}, and a sufficiently long integration time is usually required to collect a spectrum with an adequate signal-to-noise ratio (SNR). The SNR of spectra collected at different integration times shows noticeable differences, and there are also variations in spectra of the same bacterial collected by different equipment. However, the fundamental observation we made is that all the existing methods \cite{r1, r16, r17} operate under the strong assumption that the training and test data are collected under the same conditions, but the significant data distribution disparity in the training and testing spectrum collected under different acquisition conditions can result in performance degradation of existing methods. Thus, domain shift is a key challenge in developing reliable and accurate intelligent spectral identification methods, which has not been addressed in existing works.

In this paper, we propose a novel adversarial contrastive domain-generative(ACDG) learning framework for bacterial Raman spectral denoising and cross-domain identification to address the reduced recognition accuracy caused by spectral data domain differences under different acquisition conditions during model training and testing. Our core motivation is to first remove the spectral noise associated with the measure condition to preserve the most semantically meaningful spectral patterns that are invariant across varying measurement conditions. Subsequently, the denoised spectra are utilized for robust cross-domain identification.

Specifically, the ACDG framework is designed with a domain generation module and a domain task module, used respectively to generate the denoised spectra in the extended domains and to learn domain-invariant representations from both the source and extended domains. In the domain generation module, the semantic consistency-constrained style transfer is proposed to generate denoised spectra in the extended domains. Furthermore, the domain task module is designed to extract domain-invariant representations from source domain and extended domains for accurate bacterial identification. To balance the two opposite learning objectives of the effectiveness of generating denoised domains and extracting domain-invariant representations, we introduce an adversarial contrastive learning strategy to alternately optimize these two modules. Through experiments, we have demonstrated that the domain generation module can effectively remove noise from the input spectrum and enhance its SNR, while the domain task module can effectively improve the cross-domain recognition accuracy of bacterial Raman spectra under unknown acquisition conditions.

\begin{figure*}[t]
\centerline{\includegraphics[width=160mm]{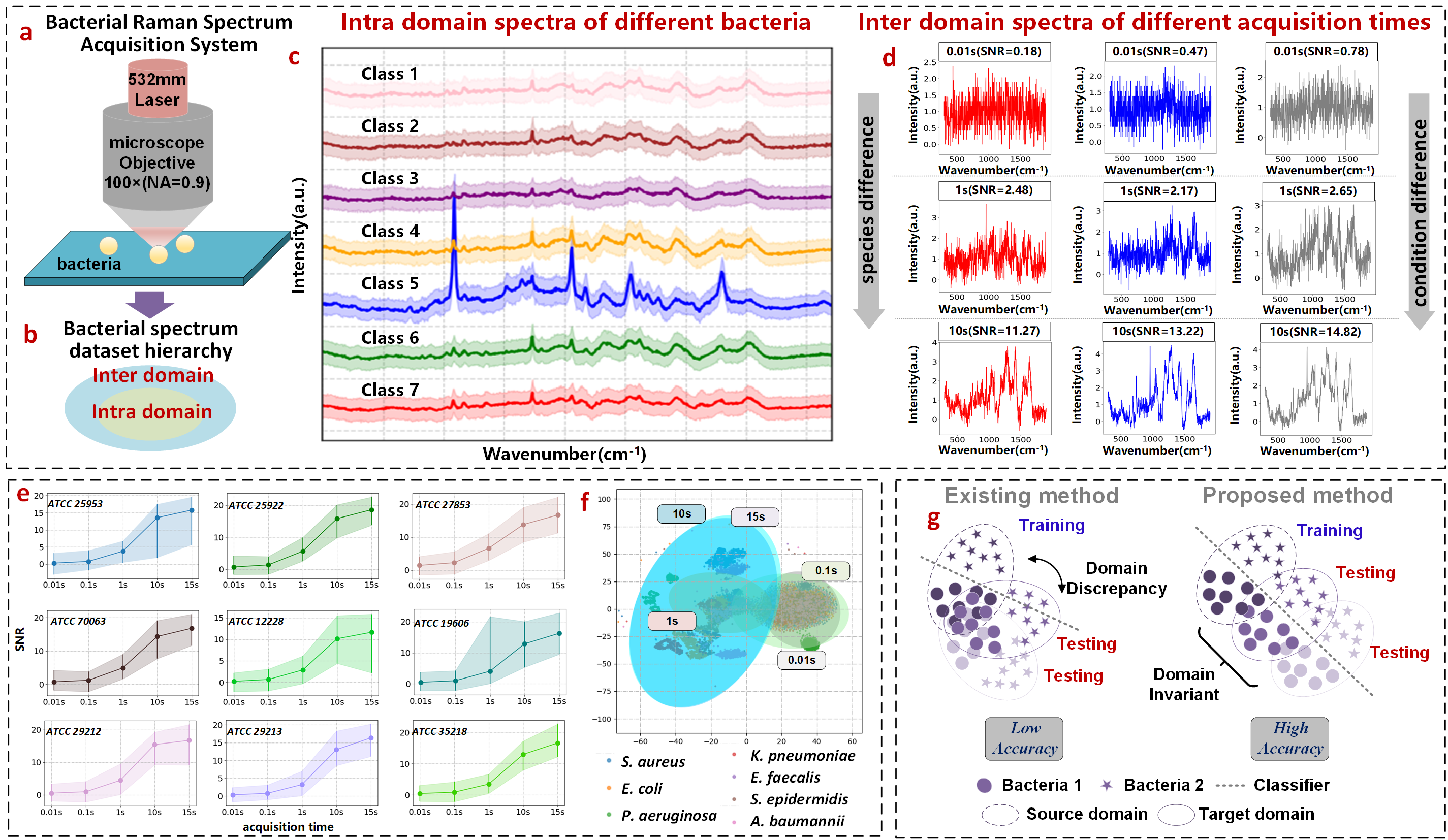}}
\caption[width=170mm]{
(a) The Raman spectrum collection setup encompasses the acquisition of Raman spectra from single bacterial cell utilizing a 532 nm excitation laser and a 100x microscope objective. 
(b) Hierarchical structure of bacterial spectral data set. (c) Intra-domain level instance: the average spectra of 7 distinct bacterial species. 
(d) Inter-domain level instance: the spectra of the same bacteria of different acquisition times. (e) The average SNR of the spectra of nine bacteria strains at different acquisition times of 0.01, 0.1, 1, 10 and 15 s. 
(f) The t-SNE visualization results.  It revealed distinct clustering patterns based on different acquisition times, resulting in five major clusters. Within each cluster, individual bacterial species formed smaller sub-clusters. 
(g) The motivation for our research. The domain discrepancies caused by different measurement conditions will directly impact the recognition performance.
}
\label{fig2}
\end{figure*}

\section{Materials and dataset}

\subsection{Bacterial spectrum dataset}

The experiments conducted in this article were carried out using the Bacterial Strains dataset\cite{r12}, which is a large-scale dataset of bacterial Raman spectra. A total of nine strains from seven species are included (as shown in Table 1), specifically \emph{Staphylococcus epidermidis} ATCC 12228, \emph{Acinetobacter baumannii} ATCC 19606, \emph{Staphylococcus aureus ATCC} 25923, \emph{Staphylococcus aureus} ATCC 29213, \emph{Enterococcus faecalis} ATCC 29212, \emph{Escherichia coli} ATCC 35218, \emph{Escherichia coli} ATCC 25922, \emph{Pseudomonas aeruginosa} ATCC 27853, and \emph{Klebsiella pneumoniae} ATCC 700603. 

All strains were cultured on tryptone soy agar at 37 °C for 24 hours. Subsequently, one colony was suspended in 5 mL of tryptone soy broth medium and cultured at 37 °C with shaking at 180 rpm for 16 hours. After reaching the stationary phase, 1 mL of the sample was taken and washed three times with sterile water. After resuspending in 1 mL of sterile water, 2 uL of each sample was deposited onto aluminum-coated slides and air-dried at room temperature. As shown in Fig. 1 (a), a Raman microscope equipped with a 532 nm laser was used to obtain single-cell Raman spectra of bacterial samples, and a 100 $\times$ objective was used to focus the laser beam onto the sample. The spectral resolution is approximately 2 cm$^{-1}$, and the wavenumber range is chosen to be 280-2186 cm$^{-1}$. For spectral measurements of each cell sample, a total of five different laser acquisition times were used, set to 0.01, 0.1, 1, 10, or 15 seconds. In the collection under these five conditions, approximately 250 cells were measured for each strain, and a total of 11,141 single-cell Raman spectra were obtained. More detailed sample preparation and measurement information can be found in \cite{r12}.

\begin{table}
\caption{Bacterial strain data set configuration}
\label{table}
\setlength{\tabcolsep}{3pt}
\centering
\begin{tabular}{p{88mm}}
${\includegraphics[width=88mm]{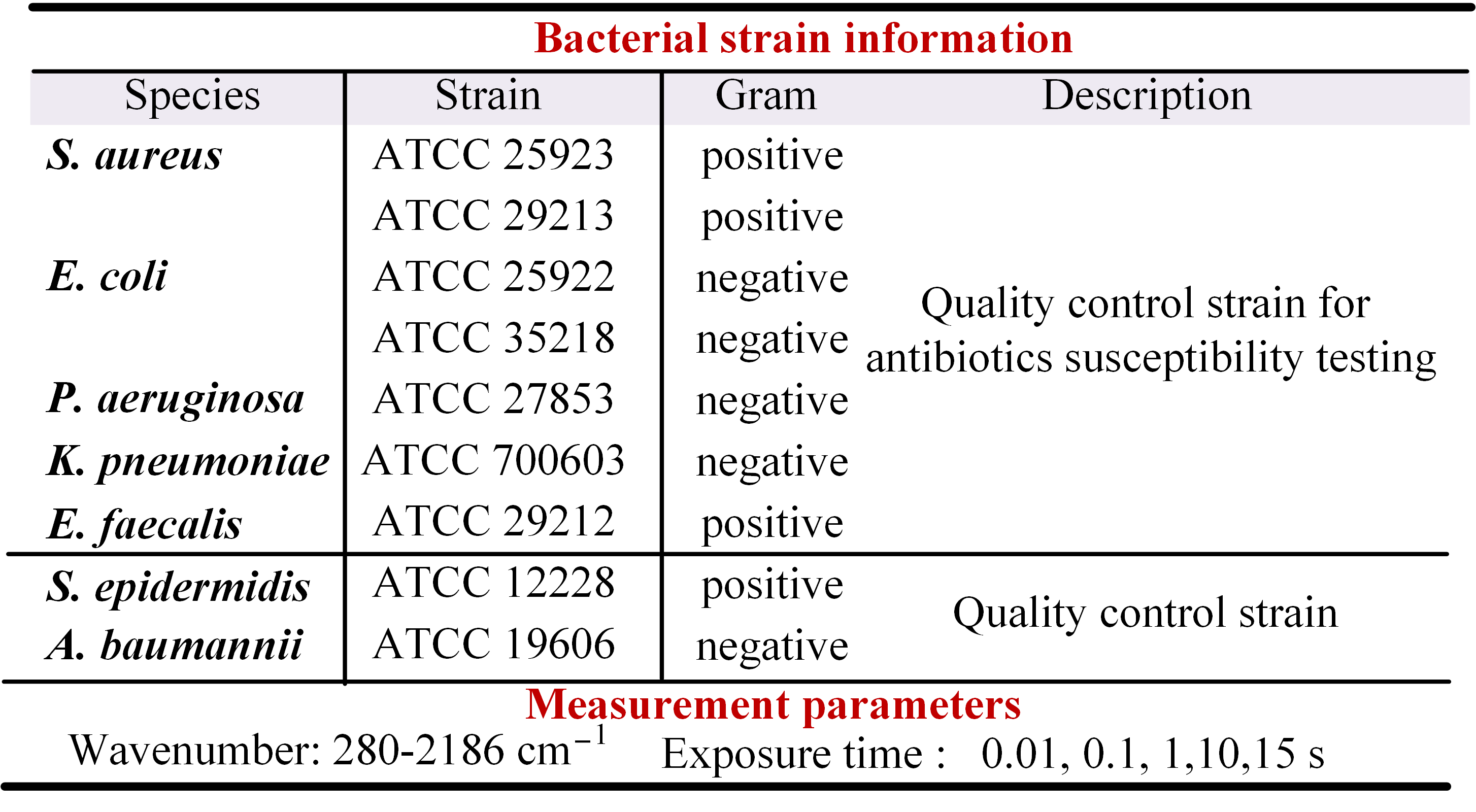}}$
\end{tabular}
\label{table_button}
\end{table}

\subsection{Spectral property analysis}

\subsubsection{Dataset hierarchy}

After acquiring single-cell spectral data, we conducted a series of analyses on their properties. Firstly, we divided the spectra into two data hierarchical levels, namely, intra-domain and inter-domain, as shown in Fig. 1 (b).

At the intra-domain level, it refers to the spectra of different species of bacteria obtained under the same measurement condition, especially the laser acquisition time. In Fig. 1 (c), we provide an example, which is the average spectrum of all spectra of 7 species of bacteria. It can be observed that the spectra of different bacterial species exhibit complex patterns, with the differences at this level mainly being species differences.

Concerning the inter-domain level, it refers to the spectra obtained under measurement conditions using different laser acquisition times. Similarly, an example is provided in Fig. 1 (d), where we show the spectra acquired for the same bacterial sample at different acquisition times. It can be seen that even for the same bacterial sample, the spectra obtained under different laser acquisition times exhibit significant differences. The shorter the integration time, the greater the spectral noise, while the longer the integration time, the more effective spectral information is obtained. The differences at this level are mainly due to condition differences.

To the best of our knowledge, all-most existing methodologies\cite{r1,r16,r17} have predominantly concentrated on addressing the spectrum identification challenge at the intra-domain level. However, concerning the identification predicament at the inter-domain level, about cross-domain recognition under different measurement conditions, this study stands as the first endeavor to address this issue.

\subsubsection{SNR analysis}

In order to quantitatively evaluate the effectiveness of bacterial spectra acquired under different laser integration times, we introduced the signal-to-noise ratio (SNR) as a quantitative index to quantify it. Specifically, for spectrum $\mathcal{X}$, its SNR is defined as:
 \begin{equation}
\mathrm{SNR}(\mathcal{X}) =10\times \mathrm{log}_{10}\left (\frac{ \frac{1}{L_S} \sum\mathcal{X}_S^{2}}{ \frac{1}{L_N} \sum\mathcal{X}_N^{2}}  \right ) 
\end{equation}
where $\mathcal{X}_S$ represents the signal segment in the spectrum, and $\mathcal{X}_n$ represents the biological noise segment in the spectrum (1800-1900 cm$^{-1}$ according to \cite{r12}). The lengths of these segments are denoted as $L_S$ for the signal and $L_N$ for the noise.

As shown in Fig. 1 (e), we calculated the SNR curve of nine strains in 11141 spectral samples as a function of laser acquisition time. We found that for each strain, its average SNR showed an overall upward trend as the acquisition time increased. At the same time, due to the individual heterogeneity of bacteria, the spectral SNR within the same strain also fluctuated to a certain extent under the same integration time. When the integration time exceeds 1 s, the average SNR improvement of the strain decreases.

\subsubsection{T-SNE visualization}

In order to more intuitively understand the hierarchical relationships in the dataset, we used the t-distributed stochastic neighbor embedding (t-SNE) algorithm\cite{r19} to perform dimensionality reduction visual analysis on 11,141 spectra. The visualization result reduced to two dimensions is shown in Fig. 1 (f). It can be found that five different measurement conditions (laser acquisition time) divide the data into five large clusters, among which the clusters with the 0.01-0.1s and 10-15s pairs integration time have a large degree of overlap. Furthermore, in each cluster, seven species form sub-clusters respectively. Since the \emph{Pseudomonas aeruginosa} ATCC 27853 specie shows strong signal reflection at specific wavelengths(749, 1128, 1312, and 1584 cm$^{-1}$), it becomes the most discriminative cluster in each measurement condition.

\subsubsection{Research motivation}
Going deeper, we can find that since these five large measurement condition clusters exhibit different data distributions, as shown in upper of Fig. 1 (g), in the existing methods, the classifier model(indicated by the gray dotted line) trained on source domain training data acquired under one measurement condition (for example, at 0.1s acquisition time), will result in degraded identification performance if directly applied to target domain testing data obtained under another measurement condition (for example, at a 1s acquisition time). The reason for this phenomenon is the domain discrepancy between the training and testing data. 

This phenomenon will have a serious impact on the actual clinical application of the intelligent recognition model for the following reason: unlike a stable laboratory environment, in actual clinical applications, various unknown factors will cause changes in measurement conditions, leading to changes in data distribution, which in turn reduce the classification performance of the model.

 As depicted in the lower of Fig. 1(g), our proposed method is designed to eliminate noise associated with measurement conditions while retaining domain-independent spectral semantic patterns. Through this approach, the model acquires domain-invariant representations, enabling the establishment of domain-shared classification boundaries and facilitating cross-domain recognition. To the best of our knowledge, this is the first work that addresses the identification of Raman spectra under varying measurement conditions.

\section{Methodology}

\subsection{Problem formulation}

Let $\mathcal{X}_i\in \mathbb{R}^{L}$ and $\mathcal{Y}_i \in \mathbb{R}^{1}$ represent the spectral sample and bacterial species label, respectively, where the $L$ is the length of the spectral data. The measurement condition domain $\mathbf{D} =\left \{ \mathbf{X} ,P(\mathbf{X} ) \right \}$ is defined as the dataset $\mathbf{X}=\left \{ \mathcal{X}_i \right \}^{N}_{i=1}$ that contains $N$ samples and its marginal distribution $P(\mathbf{X} )$. To realize the identification of bacterial spectra, the system needs to complete the task $\mathcal{T} =\left \{ \mathbf{Y}, \mathbf{C} \left ( \cdot  \right ) \right \}$ that includes label space $\mathbf{Y}=\left \{ \mathcal{Y}_i \right \}^{N}_{i=1}$ and predictive model $\mathbf{C} \left ( \cdot  \right )$.
\begin{figure*}[t]
\centerline{\includegraphics[width=170mm]{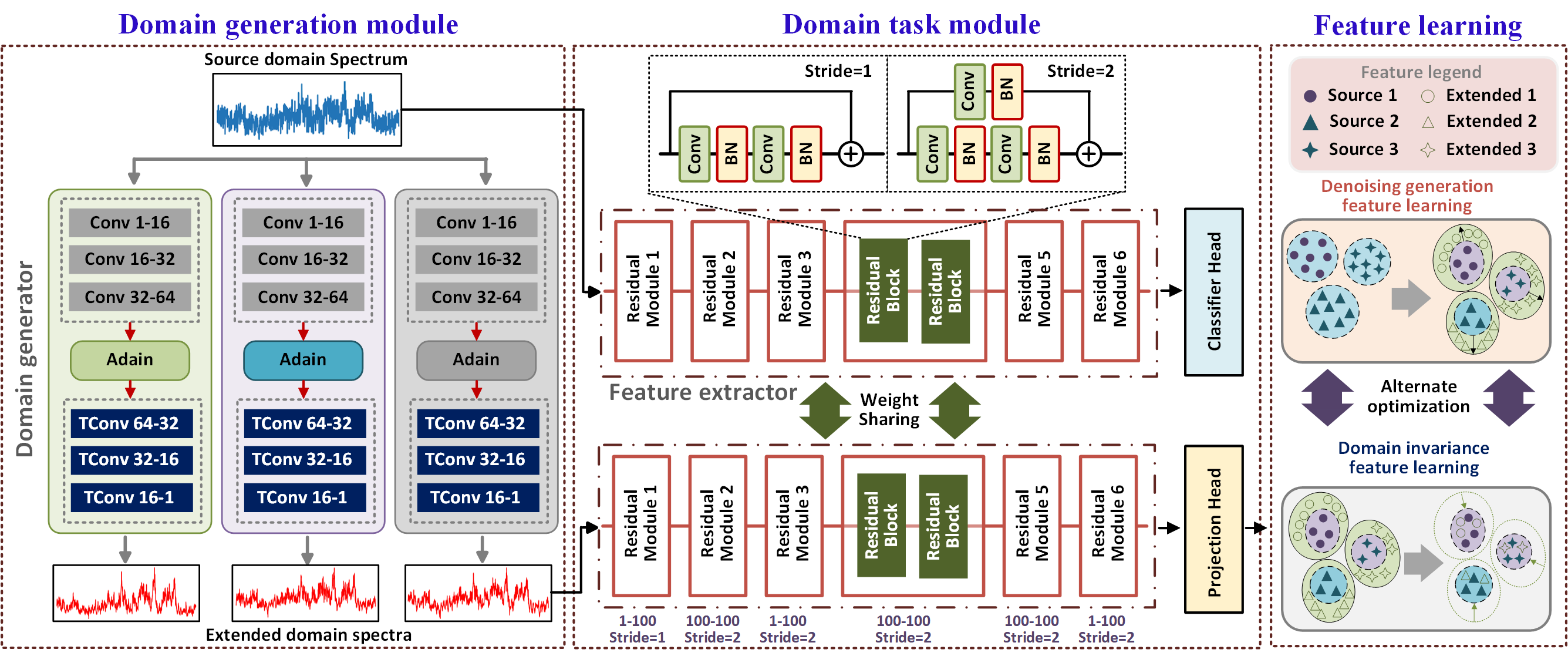}}
\caption[width=170mm]{
The architecture of the proposed ACDG. It comprises two fundamental modules: the domain generation module and the domain task module. These modules are iteratively optimized using an adversarial feature learning strategy. The domain generation module is composed of multiple domain generators $\mathcal{G}$, which are implemented through the semantic consistency-constrained style transfer network. These generators are responsible for producing denoised spectra in the extended domains. 
The domain task module comprises weight-sharing Siamese feature extractors $\mathcal{F}$, projection head $\mathcal{P}$, and classification head $\mathcal{C}$. The feature extractor $\mathcal{F}$ is structured with multiple residual modules, each comprising two residual blocks. In the first block, the stride is uniformly 1, and the number of feature channels remains unaltered. In the second block, the stride was dynamically adjusted to either 1 or 2, and the number of feature channels was also modified. The alterations in feature channels and stride sizes within each module are noted, for example, "1-100" signifies that the number of feature channels transitions from 1 to 100. The "Conv", "BN", and "TConv" represent the convolutional layer, batch normalization layer, and transpose convolutional layer, respectively. Both $\mathcal{C}$ and $\mathcal{F}$ are implemented by a basic fully connected layer. }
\label{fig2}
\end{figure*}
When considering multiple, for example, $M$ measurement condition domains $\left \{ \mathbf{D}_m \right \}^{M}_{m=1}$, the marginal distribution $P(\mathbf{X})_m$ of each measurement condition is different. Existing methods adopt the domain-task corresponding approach. Specifically, under each measurement condition $\mathbf{D}_m$, the system needs to complete domain-specific tasks $\mathcal{T}_m$. Therefore, a total of $M$ tasks need to be completed, which means that $M$ independent prediction models $\left \{ \mathbf{C} \left ( \cdot  \right )_m\right \}^{M}_{m=1}$ need to be trained on labeled datasets $L_m=(\mathbf{X}_m, \mathbf{Y}_m)=\left \{ \mathcal{X}_m^i, \mathcal{Y}_m^i  \right \}_{i=1}^{N_m}$ in their respective measurement condition domains.

In our proposed paradigm, we do not need the measurement condition dataset of all $M$ domains; instead, only one source domain $\mathbf{L}_s=(\mathbf{X}_s, \mathbf{Y}_s)=\left \{ \mathcal{X}_s^i, \mathcal{Y}_s^i  \right \}_{i=1}^{N_s}$ with the $N_s$ sample is available for training, while the other $M-1$ measurement domains are inaccessible during training, denoted as the target domains $\mathbf{D}_t$. It is worth noting that due to different measurement conditions, the data distribution of the source domain and the target domain is different, $P(\mathbf{X})_s\ne P(\mathbf{X})_t$. Therefore, in this case, the system only needs to complete one task $\mathcal{T}_s$. To address the above difficulties, we proposed the ACDG that adopts a single-domain generalization learning paradigm\cite{r20, r21}, aiming to learn a prediction model $\mathbf{C} \left ( \cdot  \right )_s$ on the training data from single source domain $\mathbf{D}_s$ that can generalize to other unknown target domains $\mathbf{D}_t$.

\subsection{ACDG overview} 
The overall framework of the proposed ACDG is shown in Fig. 2. The proposed ACDG consists of two parts: the domain generation module and the domain task module. 
In the domain generation module, we employed the domain generator $\mathcal{G}$ based on the proposed semantic consistency-constrained style transfer network, to generate different degrees of denoised spectra in the extended domain with the source domain spectra. 
In the domain task module, we introduce the feature extractor network $\mathcal{F}$, classification head network $\mathcal{C}$, and projection head network $\mathcal{P}$, aiming to extract domain-invariant features from the source domain and extended domains to achieve accurate cross-domain bacterial identification. 
These two modules implement semantic feature learning through the two adversarial goals of denoising generation and domain invariance. The details of ACDG will be expanded in subsequent subsections.

\subsection{Domain-generative denoising}

In the domain generation module, $\mathcal{G}$ is trained to generalize spectra from a single source domain $\mathbf{D}_s$ to the extended domains  $\mathbf{D}_e$:
 \begin{equation}
\mathcal{X}_{E}=\mathcal{G}\left ( \mathcal{X}_S \right ) 
\end{equation}
where $\mathcal{X}_{S}$ and  $\mathcal{X}_{E}$ represent the spectrum in the source domain and the extended domain respectively. 
As shown in Fig. 2, the $\mathcal{G}$ is implemented by the proposed semantic consistency-constrained style transfer network, which is composed of an encoder $\mathcal{E}$, a style transfer module $\mathcal{S}$, and a decoder $\mathcal{D}$. The encoder and decoder comprise stacks of convolutional layers and transpose convolutional layers, while the style transfer module consists of the Adaptive Instance Normalization(AdaIN\cite{r22}) layer. The source domain spectrum $\mathcal{X}_{S}$ passes through the encoder $\mathcal{E}$ to obtain hidden features $\mathcal{H}_{\mathcal{X}_S}=\mathcal{E}\left ( \mathcal{X}_S \right )$. Subsequently, in $\mathcal{S}$, a set of learnable mean and standard deviation parameters $\left \{ \mu_\mathcal{S},\sigma_\mathcal{S}  \right \} $ are introduced to generate extended stylized features $\mathcal{H}_{\mathcal{X}_S}^{+}$:
 \begin{equation}
\mathcal{H}_{\mathcal{X}_S}^{+}=\mathcal{S}(\mathcal{H}_{\mathcal{X}_S})=\sigma_\mathcal{S}\Big (\frac{\mathcal{H}_{\mathcal{X}_S}-\mu (\mathcal{H}_{\mathcal{X}_S})}{\sigma (\mathcal{H}_{\mathcal{X}_S})}\Big)+\mu_\mathcal{S}  
\end{equation}

Lastly, by feeding $\mathcal{H}_{\mathcal{X}_S}^{+}$ into the decoder $\mathcal{D}$, spectra in the extended domain $\mathcal{X}_{E}=\mathcal{D}(\mathcal{H}_{\mathcal{X}_s}^{+})$ can be obtained. We employ a multi-generator strategy to ensure the generation diversity of the extended domains. As shown in Fig. 2, we use three independent generators to generate a set of spectra $\left \{ \mathcal{X}_{E(k)} \right \} _{k=1}^{3}$ in the multiple extended domains. 

To ensure obtaining denoised spectra in the extended domains, we introduce two collaborative constraint conditions to optimize the domain generator $\mathcal{G}$, namely domain generation constraint and semantic consistency constraint.

\subsubsection{Domain generation constraint} As shown in Fig. 2, in the domain task module, spectra from the source domain and the extended domains are fed into the weight-shared Siamese feature extractor $\mathcal{F}$ to obtain their semantic representations, respectively. Subsequently, we further introduce the projection head $\mathcal{P}$ to project features of samples in the source and extended domains into the embedding space $\mathbf{Z}$.

Initially, our focus lies in prompting the domain generator $\mathcal{G}$ to procure samples within the extended domain that deviate from the distribution observed in the source domain. This objective is realized through the optimization of feature distribution within the embedding space $\mathbf{Z}$. Drawing inspiration from contrastive learning principles\cite{r23}, we firstly introduce the domain expansion constraint to learn class-domain specific representations:
 \begin{equation}
\mathcal{L} _{expansion} =-\sum_{i=1}^{N_S} \sum_{p\in P(i)}^{}\mathrm{log} \frac{ e^{ \left [ \mathcal{S}\left (  \mathcal{Z}_{S}^{i},  \mathcal{Z}_{S}^{p}\right )   /\tau  \right ] } }{\sum_{a\in A(i)} e ^{\left[ \mathcal{S}\left (  \mathcal{Z}_{S}^{i},  \mathcal{Z}_{\Omega }^{a}\right ) /\tau \right ] } }
\end{equation}
where the $\mathcal{Z}_{S}$ and $\mathcal{Z}_{E}$ represents the embedded features in $\mathbf{Z}$ of the source and extended samples. $\mathcal{S}$ is the feature similarity function implemented by the cosine similarity\cite{r24}. $\tau$ denotes the temperature coefficient which is set to 0.2. Domain set $\Omega =\left \{\mathcal{S}, E(k) \right \}$ includes the source domain $\mathcal{S}$ and all extended domains $E$, $P(i)$ and $A(i)$ indicate the positive and complete set of $i$-th embedded feature within the data batch, respectively. The positive samples in $P(i)$ are defined as sharing the same spectral category with the $i$-th samples:
\begin{equation}
P(i)=\left \{ p\in A(i):\mathcal{Y}_p=\mathcal{Y}_i  \right \} 
\end{equation}

The domain expand loss encourages the features of samples belonging to the same spectral class within the source domain to be close to each other, while the features of samples from different spectral classes or different domains are pushed apart. Therefore, $\mathcal{G}$ can be directed to generate extended spectral domains with distribution discrepancies from the source domain. As described before, we introduced the multiple-domain expansion strategy to ensure the diversity of the generated domain. Therefore, we further proposed a diversity constraint:
 \begin{equation}
\mathcal{L} _{diversity}=-\sum_{i=1}^{N_S}\sum_{k=1}^{3}\sum_{p\in P(i)}^{}\mathrm{log} \frac{ e^{ \left [ \mathcal{S}\left (  \mathcal{Z}_{E(k)}^{i},  \mathcal{Z}_{E(k)}^{p}\right )   /\tau  \right ] } }{\sum_{a\in A(i)} e ^{\left[ \mathcal{S}\left (  \mathcal{Z}_{E(k)}^{i},  \mathcal{Z}_{E}^{a}\right ) /\tau  \right ] } } 
\end{equation}

With the diversity constraint, the features of samples within the same spectral category in a given extended domain are encouraged to converge, while the features of samples from disparate spectral categories or different extended domains are simultaneously driven to diverge. Combining the above two constraints, the domain generation constraint can be obtained as:
 \begin{equation}
\mathcal{L}_{generation}=\mathcal{L}_{expansion}+\mathcal{L}_{diversity}
\end{equation}

\subsubsection{Semantic consistency constraint}

It is worth noting that the $\mathcal{G}$ may produce semantically confusing samples when generating samples with domain differences. Therefore, the optimization of $\mathcal{G}$ needs to consider the alignment of semantic information to ensure that the generated samples retain the category semantics of the source domain spectra. Therefore, we further introduce the semantic consistency constraint.

As depicted in Fig. 2, the semantic representation obtained through the feature extractor $\mathcal{F}$ is forwarded to the classification head $\mathcal{C}$ to obtain its prediction of the spectral category. The semantic consistency constraint is defined as the cross-entropy loss between the category prediction output by the classification head $\mathcal{C}$ and the true label of the generated extended domain samples:
 \begin{equation}
\mathcal{L} _{semantics}=-\sum_{i=1}^{N_S}\sum_{k=1}^{3}\mathcal{L}_{CE}(\mathcal{C}(\mathcal{F}(\mathcal{X}_{E(k)}^i)),\mathcal{Y}_s^i  )
\end{equation}
where the $\mathcal{L}_{CE}$ is the cross-entropy loss function. The aforementioned domain generation and semantic consistency constraints jointly control the optimization direction of the domain generator $\mathcal{G}$. On the one hand, the domain generation loss drives $\mathcal{G}$ to generate spectra in the extended domains that differ from those in the source domain distribution. On the other hand, the semantic consistency constraint ensures that the generated spectra in the extended domain have a distribution conducive to identification, meaning they retain the most semantically informative parts while removing irrelevant noise components. Therefore, the domain generator serves as a \textbf{spectral denoiser}. It is worth noting that unlike existing supervised denoising methods\cite{r7, r15}, our denoising approach does not require any noise-free ground truth for training; only spectral category labels are needed, offering significant advantages.

\subsection{Domain-invariant recognition}
In the domain task module, the classification task constraint is minimized to ensure that both the feature extractor $\mathcal{F}$ and the classification head $\mathcal{C}$ can accurately recognize the spectral classes of samples from both the single source domain and the extended domains:
 \begin{equation}
\mathcal{L} _{task}=-\sum_{i=1}^{N_S}\mathcal{L}_{CE}(\mathcal{C}(\mathcal{F}(\mathcal{X}_{\Omega }^i)),\mathcal{Y}_s^i)
\end{equation}

Subsequently, to attain robust cross-domain spectral recognition, the feature extractor $\mathcal{F}$ is encouraged to acquire domain-invariant representations. Thus, within the embedding space $\mathbf{Z}$, we further introduce domain invariance constraint to achieve the above objectives:
 \begin{equation}
\mathcal{L} _{invariance} =-\sum_{i=1}^{N_S} \sum_{p\in P(i)}^{}\mathrm{log} \frac{ e^{ \left [ \mathcal{S}\left (  \mathcal{Z}_{\Omega }^{i},  \mathcal{Z}_{\Omega }^{p}\right )   /\tau  \right ] } }{\sum_{a\in A(i)} e ^{\left[ \mathcal{S}\left (  \mathcal{Z}_{\Omega }^{i},  \mathcal{Z}_{\Omega }^{a}\right ) /\tau  \right ] } }
\end{equation}

By imposing the domain invariance constraint, sample features within the same spectral categories across all domains are encouraged to exhibit proximity, whereas sample features from different spectral categories generally diverge. This approach facilitates the learning of domain-independent features.

\subsection{Model training and inference}

In AGL, due to the opposite optimization directions of the domain generation and domain task modules, an adversarial alternating iterative optimization method is used to obtain optimal AGL parameters. The first step involves optimizing the feature extractor $\mathcal{F}$, projection head $\mathcal{P}$, and classification head $\mathcal{C}$ in the domain task module through the  domain task loss $\mathcal{L}_{DT}$ obtained by combining the classification task constraint $\mathcal{L} _{task}$ and the domain invariance constraint $\mathcal{L} _{invariance}$:

 \begin{equation}
\underset{\boldsymbol{\theta}_\mathcal{F}, \boldsymbol{\theta}_\mathcal{P}, \boldsymbol{\theta}_\mathcal{C}}{\mathrm{min}}\mathcal{L}_{DT}=\beta\mathcal{L}_{task}+\mathcal{L}_{invariance}   
\end{equation}
where $\beta$ is the weight coefficient balancing the contributions of the two loss terms. Subsequently, the domain generator $\mathcal{G}$ in the domain generation module is optimized through the domain generation loss $\mathcal{L}_{DG}$ that combines the domain generation constraint $\mathcal{L}_{generation}$ and semantic consistency constraint  $\mathcal{L} _{semantics}$:
 \begin{equation}
\underset{\boldsymbol{\theta}_\mathcal{G}}{\mathrm{min}}\mathcal{L}_{DG}= \alpha \mathcal{L}_{generation}  +\mathcal{L}_{semantics} 
\end{equation}
where $\alpha$ is the weight coefficient. During each training epoch, the optimizer updates the parameters of each network in the domain task module as follows:
 \begin{equation}
\begin{aligned}
&\boldsymbol{\theta}_\mathcal{F}^{e+1}\gets \boldsymbol{\theta}_\mathcal{F}^{e}-\mu \Big( \beta \frac{\partial\mathcal{L}_{task}}{\partial \boldsymbol{\theta}_\mathcal{F}^{e}} +  \frac{\partial\mathcal{L}_{invariance}}{\partial \boldsymbol{\theta}_\mathcal{F}^{e}} \Big)\\
&\boldsymbol{\theta}_\mathcal{C}^{e+1}\gets \boldsymbol{\theta}_\mathcal{C}^{e}-\mu \Big( \beta \frac{\partial\mathcal{L}_{task}}{\partial \boldsymbol{\theta}_\mathcal{C}^{e}} \Big)\\
&\boldsymbol{\theta}_\mathcal{P}^{e+1}\gets \boldsymbol{\theta}_\mathcal{P}^{e}-\mu \Big( \beta \frac{\partial\mathcal{L}_{invariance}}{\partial \boldsymbol{\theta}_\mathcal{P}^{e}} \Big)
\end{aligned}
\end{equation}
where $e$ represents the $e$-th training epoch, and $\mu$ represents the learning rate. Similarly, the optimization of parameters in the domain generation module is as follows:
 \begin{equation}
\boldsymbol{\theta}_\mathcal{G}^{e+1}\gets \boldsymbol{\theta}_\mathcal{G}^{e}-\mu \Big( \alpha \frac{\partial\mathcal{L}_{generation}}{\partial \boldsymbol{\theta}_\mathcal{G}^{e}} +\frac{\partial\mathcal{L}_{semantics}}{\partial \boldsymbol{\theta}_\mathcal{G}^{e}}\Big)
\end{equation}

\begin{algorithm}[t]
\caption{ACDG optimization method}
 \SetKwInOut{Input}{input}
\SetKwInOut{KwOut}{Models}
\KwIn{Source domain dataset $L_s=\left \{ \mathcal{X}_s^i, \mathcal{Y}_s^i  \right \}_{i=1}^{N_s}$}
 \KwOut{domain generator $\mathcal{G}$, feature extractor $\mathcal{F}$,  classification head  $\mathcal{C}$, and projection head $\mathcal{P}$}

\For{$epoch=1,...,{epoches}$}
{
Randomly extract data batches from $L_s$\;

\textbf{(a)}. \emph{domain task module training}\;
Freeze the model parameters $\boldsymbol{\theta}_\mathcal{G}$\;
Calculate loss: $\mathcal{L}_{DT}\gets \mathrm{Eq.(11)}$\;
Backpropagation to update $\boldsymbol{\theta}_\mathcal{F}, \boldsymbol{\theta}_\mathcal{P}, \boldsymbol{\theta}_\mathcal{C}$ by Eq. (13)\;

\textbf{(b)}. \emph{domain generation module training}\;
Freeze the model parameters $\boldsymbol{\theta}_\mathcal{F}, \boldsymbol{\theta}_\mathcal{P}, \boldsymbol{\theta}_\mathcal{C}$\;
Calculate loss: $\mathcal{L}_{DG}\gets \mathrm{Eq.(12)}$\;
Backpropagation to update $\boldsymbol{\theta}_\mathcal{G}$ by Eq. (14)\;
}
return $\boldsymbol{\theta}_\mathcal{F}, \boldsymbol{\theta}_\mathcal{P}, \boldsymbol{\theta}_\mathcal{C}, \boldsymbol{\theta}_\mathcal{G}$
\end{algorithm}

In general, the adversarial alternating iterative optimization method is presented in Algorithm 1. After completing the AGL training, we employ $\mathcal{G}$ to denoise the input spectrum. Subsequently, the denoised spectra from multiple generators are averaged to derive the final denoised spectrum, which is then fed into the feature extractor $\mathcal{F}$ and classification head $\mathcal{C}$ for spectrum identification.

\section{Experimental results}

In this section, we will evaluate the performance of the proposed ACDG framework, focusing primarily on spectral denoising and spectral identification tasks.

\subsection{Experiment implementation}

\begin{table}
\caption{Task information}
\label{table}
\setlength{\tabcolsep}{3pt}
\centering
\begin{tabular}{p{88mm}}
${\includegraphics[width=88mm]{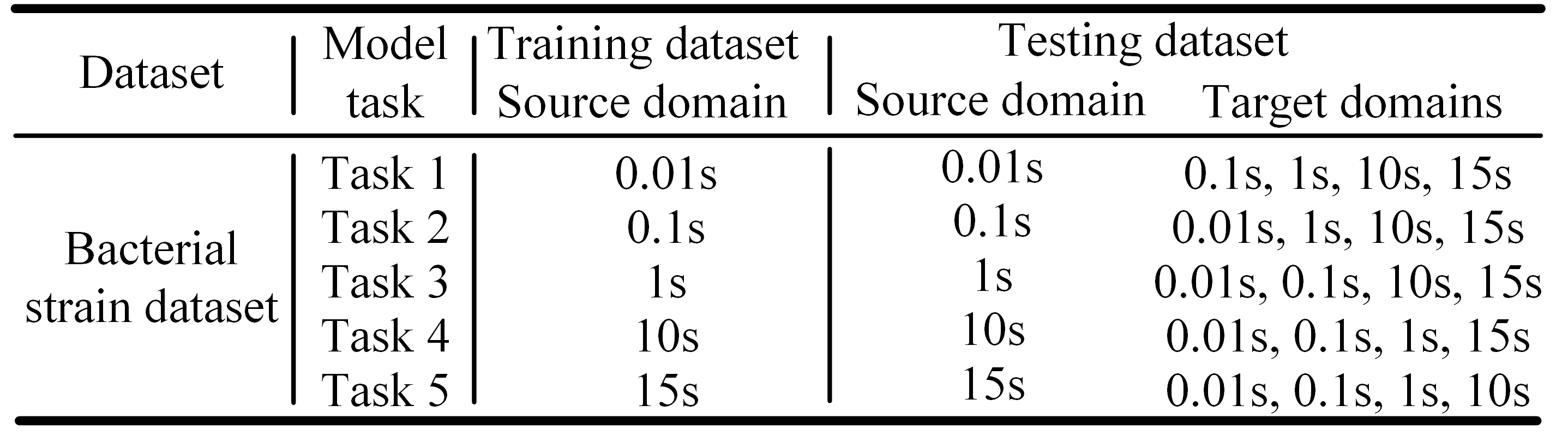}}$
\end{tabular}
\label{table_button}
\end{table}
\subsubsection{Task settings}

In our experiments, detailed in Section II, we partitioned the dataset into five subsets corresponding to distinct measurement conditions, specifically five different sample  acquisition times. For each subset, we defined tasks as outlined in Table 2. Each task's training set comprised 20$\%$ of the source domain data under a single measurement condition, whereas the test set encompassed all data from the remaining four measurement conditions along with the remaining 80$\%$ of the source domain data.

Hence, this task configuration allows for the assessment of the model's capabilities in both intra-domain and inter-domain recognition. Moreover, the varying acquisition times of training data across different tasks serve to further elucidate the influence of training data at different acquisition times on the model's performance.

\subsubsection{Implementation details} 
We trained the ACDG from scratch using the AdamW optimizer, with a learning rate of 1e-4, a batch size of 12, and for 100 epochs. All the experiments were conducted on a computer equipped with Xeon(R) Gold 6226R CPUs@2.90GHz and two NVIDIA A100 GPUs with 40GB of memory.

\subsubsection{Comparative baselines} 
ACDG consists of two modules, one dedicated to spectrum denoising and the other to spectrum identification. Hence, we will compare it with existing methods based on these two aspects. 

For spectral denoising, our method does not require real noise-free spectra as ground truth for model training, thus it is considered an unsupervised denoising method. Therefore, we compare it with existing unsupervised denoising methods. These methods include  wavelets-based denoising method\cite{r25}, principal component analysis (PCA)-based denoising method\cite{r27}, Savitzky-Golay denoising method\cite{r28}, Wiener filtering denoising method\cite{r29}, and autoencoder-based denoising method\cite{r26}. 

For spectral recognition, we consider both machine learning-based and deep learning-based methods for comparison. Among them, machine learning-based methods include quadratic discriminant analysis (QCA)\cite{r30}, decision trees (DT)\cite{r31}, and support vector machines (SVM)\cite{r32}. Methods based on deep learning include various deep neural networks, such as artificial neural networks (ANN)\cite{r33}, convolutional neural networks (CNN)\cite{r34}, long short-term memory network (LSTM)\cite{r35}, Gram angle field with 2D convolutional neural network (GAF+2DCNN)\cite{r36}, residual network (ResNet)\cite{r1}, and scale-adaptive network (SANet) \cite{r16}. Among these methods, ResNet and SANet represent state-of-the-art (SOTA) techniques.

\subsubsection{Evaluation indicators} 

For both denoising and recognition tasks, we employ quantitative evaluation metrics to gauge the model's performance. Specifically, we use SNR to assess denoising effectiveness. For recognition tasks, we employ a composite of performance metrics, including accuracy, precision, specificity, recall, and the F1 score.

\begin{figure*}[t]
\centerline{\includegraphics[width=170mm]{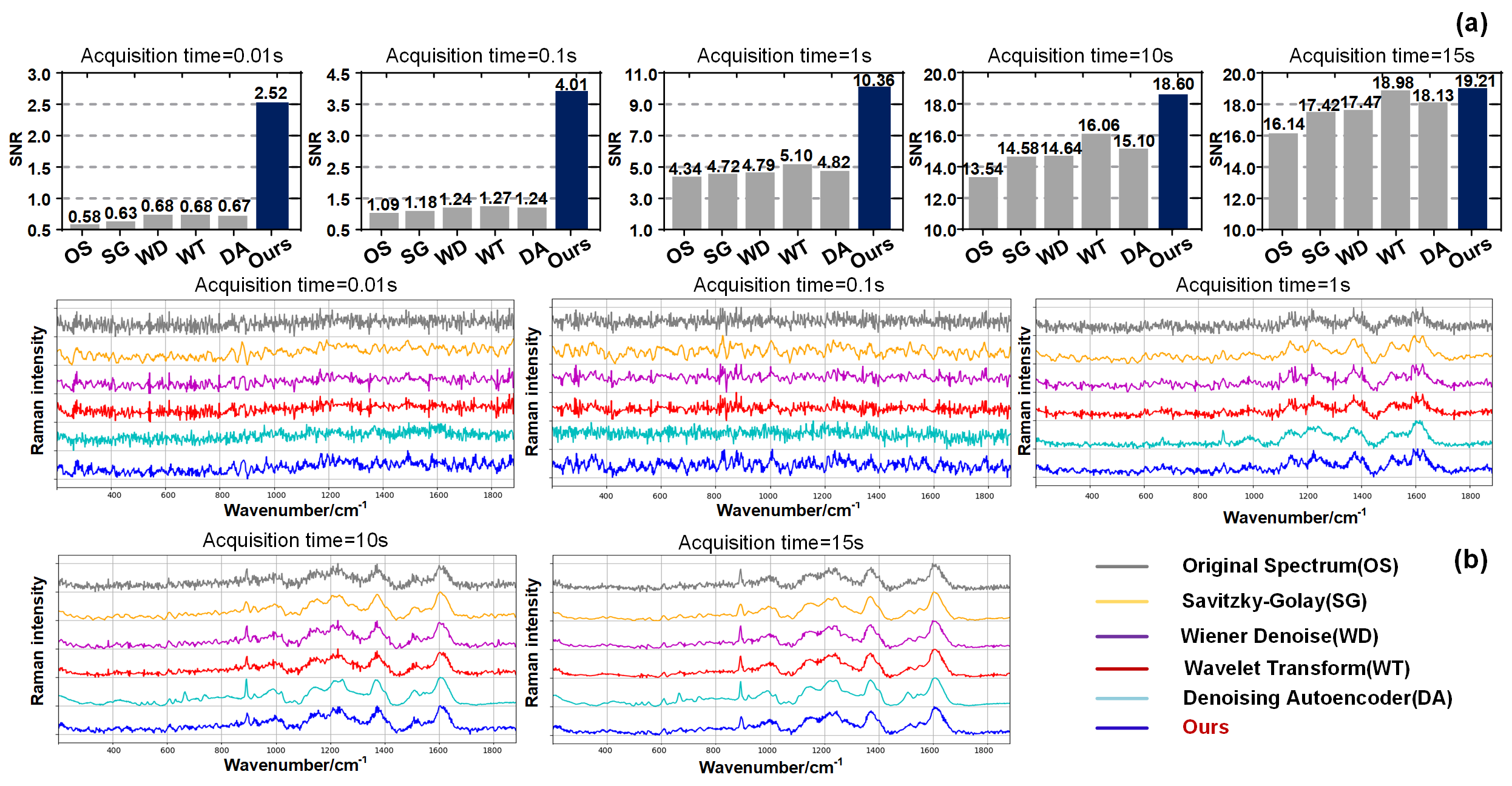}}
\caption[width=170mm]{
Experimental outcomes concerning spectral denoising. (a) Quantitative comparison of spectral SNR across different acquisition times employing different comparative methods. (b) Qualitative comparisons of the denoised spectra yielded by diverse comparative methods across varying acquisition times.
}
\label{fig2}
\end{figure*}

\subsection{Denoising experimental results} 

In this section, a series of experiments is conducted to compare the spectral denoising efficacy of different methodologies, with the results depicted in Fig. 3.

First,the quantitative evaluation of the average SNR across all spectra  across five acquisition times for the compared methods are presented in Fig. 3 (a), from which we can obtain the following observations: \textbf{\emph{1}})~To obtain original spectra with a high average SNR, a long acquisition time is usually required.  \textbf{\emph{2}})~While other established methodologies may enhance the SNR of the original spectrum to some degree, the benefits are not pronounced, particularly at shorter acquisition times such as 0.01s, 0.1s, and 1s. The augmentation in SNR achieved through these approaches is less than 1.0. \textbf{\emph{3}})~ The denoising efficacy of the proposed ACDG is noteworthy. Across various acquisition times, the denoised spectrum acquired via ACDG consistently attains the highest SNR. Specifically, under five distinct acquisition time, the average SNR improvements achieved are +1.94, +2.92, +6.02, +5.06, and +3.07, respectively. Moreover, the SNR of spectra acquired at acquisition times of 0.01s, 0.1s, and 1s exhibited an increase from 0.58 to 2.52, from 1.09 to 4.01, and from 13.54 to 18.60, respectively, following the application of ACDG denoising. Consequently, employing ACDG enables a reduction in data acquisition time by around one order of magnitude(\textit{e.g.} from 1s to 0.1s),  suggesting that ACDG holds the potential to considerably accelerate Raman spectroscopy analysis,

Furthermore, in Fig. 3 (b), we present a qualitative assessment of the denoising effectiveness of various methods across varying acquisition times. The results revealed that the visual fidelity of original Raman spectra at short acquisition times (\textit{e.g.}, 0.01s and 0.1s) was in inferior quality. Specifically, the Raman peak was overwhelmed by noise, leading to an inability to discern clear spectral patterns. Conversely, following ACDG denoising, clearer spectral features can be observed, which is beneficial to the subsequent biometric identification.

\begin{table*}
\caption{Comprehensive comparison results of different methods for intra- and inter-domain spectral identification}
\label{table}
\setlength{\tabcolsep}{3pt}
\centering
\begin{tabular}{p{160mm}}
${\includegraphics[width=160mm]{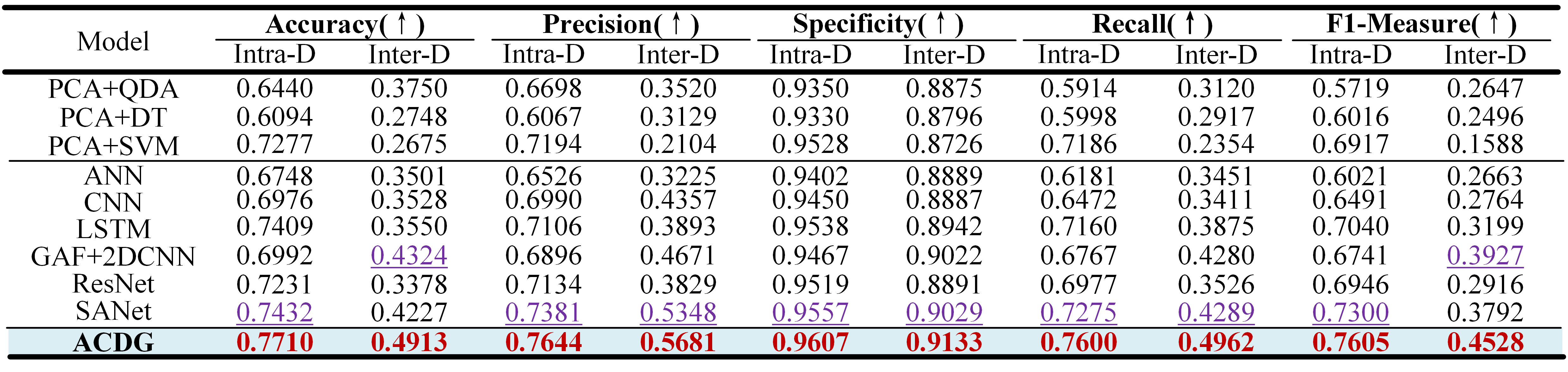}}$
\end{tabular}
\label{table_button}
\end{table*}

\subsection{Identification experimental results} 
In this section, we conduct a comparative analysis of various methods for Raman spectral identification. The efficacy of recognition can be evaluated from two primary perspectives: intra-domain and inter-domain recognition performance.

\subsubsection{Comprehensive results} 

First, we conducted a comprehensive comparison of the recognition performance among various methods, with the outcomes summarized in Table 3. The results indicate that both traditional machine learning approaches and contemporary deep learning techniques exhibit higher intra-domain recognition accuracy compared to inter-domain ones. This phenomenon can be attributed to the discrepancy in data distribution encountered in inter-domain recognition tasks. Furthermore, it is noteworthy that deep learning-based methodologies generally outperform those based on traditional machine learning. This superiority stems from the inherent capability of deep neural networks to automatically extract more discriminative spectral pattern features. 

Furthermore, in comparison to state-of-the-art (SOTA) methods such as ResNet and SANet within the realm of spectral recognition, our approach demonstrates superior inter-domain and intra-domain recognition performance. This superiority can be attributed to the underlying assumption of existing deep learning methods, including ResNet and SANet, which rely on the premise that training and test data share the same distributions. However, in practical scenarios, distribution shifts occur due to variations in measurement conditions. Consequently, these conventional methods are unable to achieve robust recognition performance, particularly in inter-domain recognition tasks.

\begin{table*}
\caption{Task-level fine-grained comparison results of recent SOTA methods for intra-domain spectral identification}
\label{table}
\setlength{\tabcolsep}{3pt}
\centering
\begin{tabular}{p{160mm}}
${\includegraphics[width=160mm]{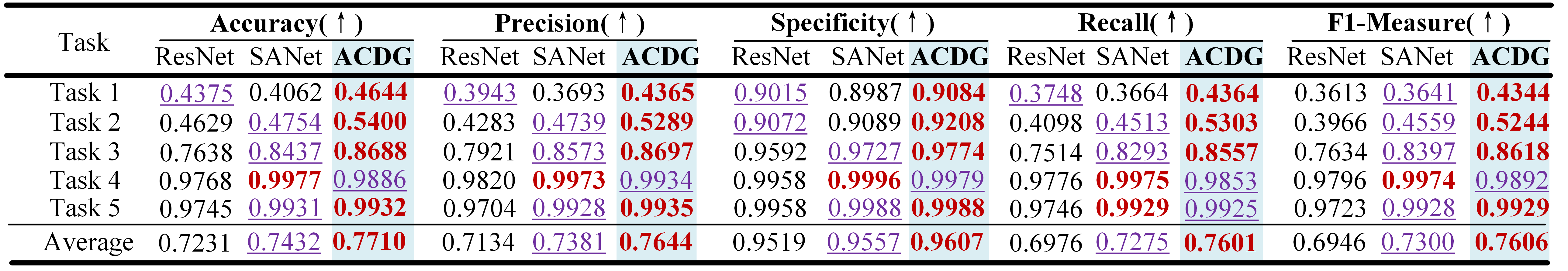}}$
\end{tabular}
\label{table_button}
\end{table*}

\begin{table*}
\caption{Task-level fine-grained comparison results of recent SOTA methods for inter-domain spectral identification}
\label{table}
\setlength{\tabcolsep}{3pt}
\centering
\begin{tabular}{p{160mm}}
${\includegraphics[width=160mm]{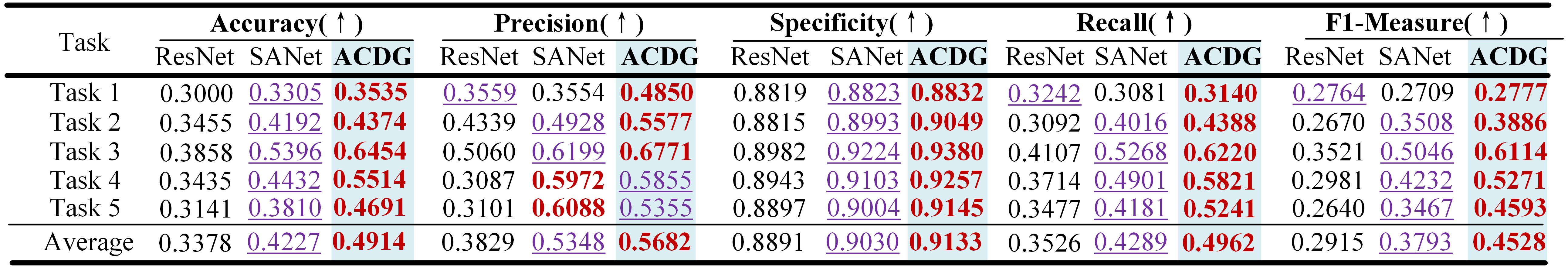}}$
\end{tabular}
\label{table_button}
\end{table*}

Conversely, our ACDG framework excels in extracting maximal semantic information via the domain generation module while effectively filtering out irrelevant noise signals. Consequently, this approach alleviates the adverse effects of noise on recognition accuracy and substantially enhances both intra-domain and inter-domain recognition capabilities. Specifically, in the realm of intra-domain recognition, our ACDG surpasses the sub-optimal SANet by achieving a noteworthy improvement of +2.78$\%$, +2.63$\%$, +0.50$\%$, +3.25$\%$, and +3.05$\%$ in accuracy, precision, Specificity, recall, and F1 score. Similarly, in the context of inter-domain recognition, the enhancement reaches significant +6.86$\%$, +3.33$\%$, +1.04$\%$, +6.73$\%$, and +5.56$\%$.

\begin{figure*}[t]
\centerline{\includegraphics[width=160mm]{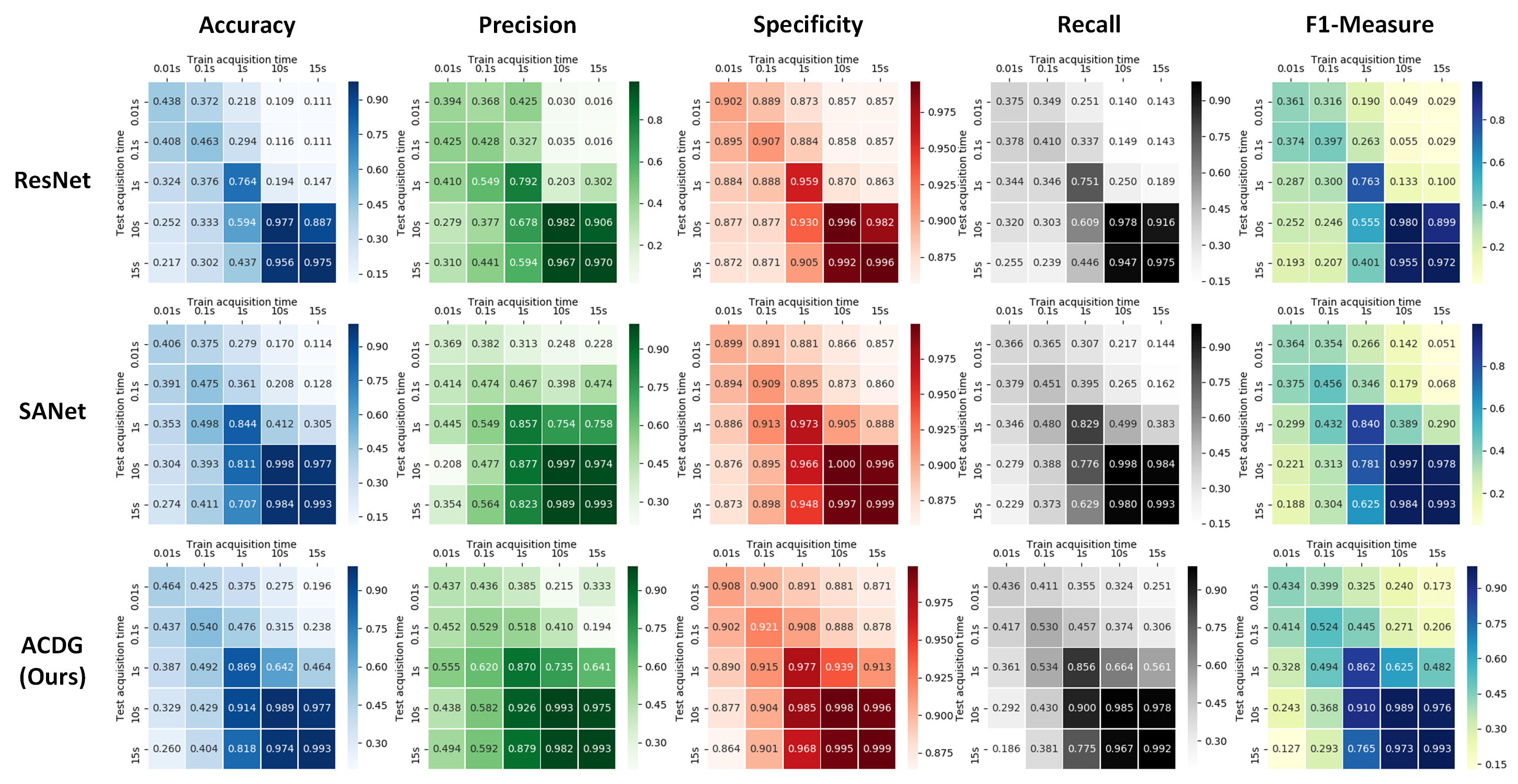}}
\caption[width=160mm]{
Heat maps of detailed quantitative recognition results comparing methods ResNet, SANet, and the proposed ACDG.
}
\label{fig2}
\end{figure*}

\subsubsection{Fine-grained task-level results}

Furthermore, we present a detailed analysis of task-level intra-domain and inter-domain fine-grained recognition outcomes in comparison with ResNet and SANet, as delineated in Tables 4 and 5. Additionally, we visually represent the detailed outcomes through heat maps, depicted in Fig. 4.

The outcomes underscore the superior performance of our ACDG across nearly all tasks. Additionally, the following observations were noted: \textbf{\emph{1}})~ While our method excels in both intra- and inter-domain recognition tasks, it exhibits greater gains in the context of inter-domain recognition. This can be attributed to the domain generation module's effectiveness in eliminating noise from diverse measurement conditions. Moreover, the domain task module aids in extracting domain-invariant features, thereby collaboratively enhancing recognition accuracy for cross-domain spectra. \textbf{\emph{2}})~In the context of intra-domain recognition tasks, our method exhibits a greater performance improvement in tasks characterized by shorter training acquisition times (\textit{e.g.}, Task 1 and Task 2). This observation is attributed to the higher intensity of noise present in the testing spectra for these tasks. Our ACDG effectively mitigates this type of noise, resulting in enhanced performance. Conversely, for tasks with longer training acquisition times (\textit{e.g.}, Tasks 4 and 5), the noise levels in the testing spectra are already relatively low. As a result, the gains achieved by noise removal are less significant. \textbf{\emph{3}})~Conversely, within the realm of inter-domain recognition tasks, our method demonstrates a greater performance gain in tasks characterized by longer training acquisition times (such as Tasks 4 and 5). This phenomenon is attributed to the shorter acquisition time of the testing spectra in these tasks, which leads to a higher noise level compared to that during training. For instance, in Task 5, the training acquisition time is 15s, while the testing acquisition time ranges from 0.01s to 10s. Under such conditions, traditional methods are susceptible to noise interference, leading to decreased accuracy. In contrast, our method effectively eliminates these noises, resulting in a more robust recognition performance.

\subsection{Discussion} 

\subsubsection{Efficiency of training acquisition times} 

Achieving intelligent spectral identification necessitates the utilization of a substantial quantity of spectral samples for model training. The duration required to gather training samples is directly proportional to the acquisition time of each spectrum. Therefore, choosing the shortest training acquisition time can accelerate the entire acquisition process. In this section, we delve into the influence of training acquisition time on the overall recognition performance.

The results are depicted in Fig. 5, illustrating the average recognition performance across inter-domain and intra-domain scenarios for ResNet, SANet, and our proposed ACDG under varying training acquisition times.

\begin{figure*}[t]
\centerline{\includegraphics[width=160mm]{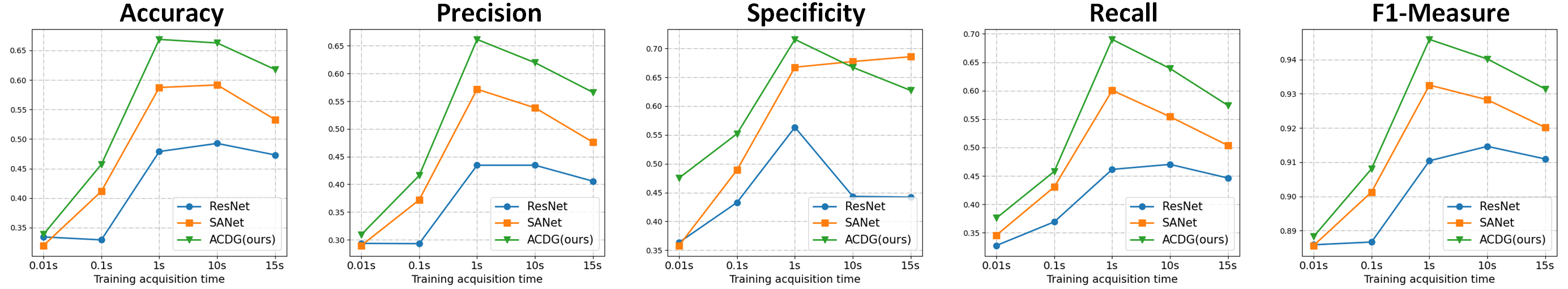}}
\caption[width=160mm]{
Overall recognition performance of the ACDG and comparison methods ResNet and SANet under different training acquisition times.
}
\label{fig2}
\end{figure*}

\begin{figure*}[t]
\centerline{\includegraphics[width=160mm]{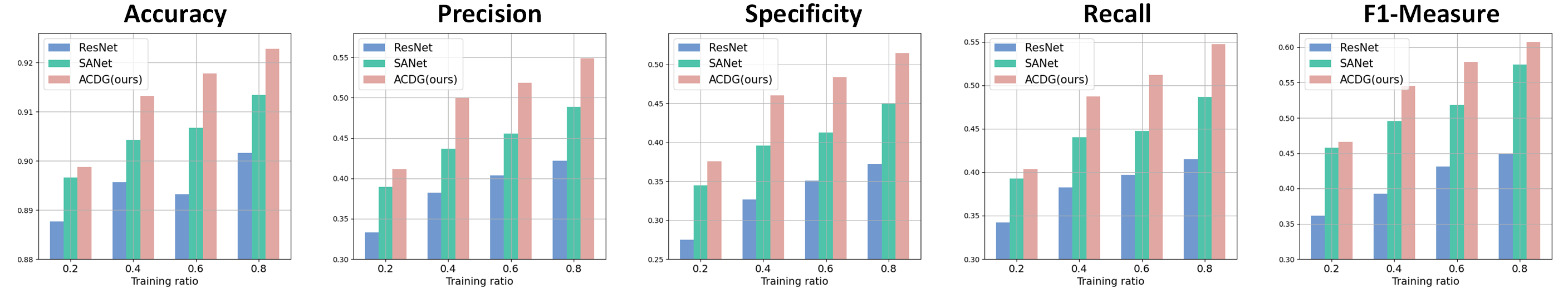}}
\caption[width=160mm]{
Overall recognition performance of the ACDG and comparison methods ResNet and SANet under different training spectral ratios.
}
\label{fig2}
\end{figure*}

The results indicate a trend where the recognition performance of various methods firstly rises and then declines as the training acquisition time increases. This pattern arises due to several factors. Firstly, when the training time is excessively short, the SNR of the training spectrum becomes too low, hindering the model's ability to learn meaningful spectral patterns. Conversely, excessively long training times result in high SNR for the training spectral data. However, this leads to significant domain disparities with the low-SNR spectral data present in the testset. For instance, a model trained using spectra with a 15s acquisition time may face challenges when tested with spectra acquired in just 0.01s, consequently causing a decline in test performance. 

Moreover, the outcomes also demonstrate that regardless of the training time employed, our proposed ACDG method consistently attains optimal recognition accuracy. This underscores the robustness of ACDG, as it consistently delivers optimal performance regardless of variations in training acquisition time.

\subsubsection{Influence of training sample numbers}

Another determinant directly impacting training costs is the quantity of training samples. Collecting a substantial array of spectral samples necessitates considerable human effort. Thus, it is imperative to explore the model's performance across various quantities of training spectral samples. In this section, we considered varying proportions of training samples, specifically 0.2, 0.4, 0.6, and 0.8, to examine the average the model recognition performance across cross-domain and intra-domain scenarios under these conditions. The outcomes are detailed in Fig. 6.

As the number of training samples decreases, both ResNet and SANet exhibit degraded performance due to the overfitting phenomenon associated with small sample sizes. Conversely, our proposed ACDG consistently achieves optimal performance across various proportions of training data. The rationale behind this lies in the domain generative denoising module, which expands the data range available to the domain task module by creating new samples within the extended domains. This mechanism significantly mitigates the overfitting phenomenon. These results showcase the robustness of this method when confronted with scenarios involving limited samples.

\subsubsection{Trade-off of hyper-parameters} 

Hyper-parameters play a significant role in optimizing the model during training. Within the proposed ACDG framework, the pivotal hyperparameters are the weight coefficients $\beta$ and $\alpha$ in the loss function, as defined by Eq. (11) and (12). To investigate their influence on both cross-domain and intra-domain recognition performance, this section undertakes a grid search and sensitivity analysis on these hyper-parameters . The findings from this analysis are presented in Fig. 7. The search range for $\beta$ and $\alpha$ is specified between 0.001 and 1, with increments of tenfold. The outcomes reveal that when $\beta$ is set to 0.1 and $\alpha$ to 0.01, the ACDG achieves optimal average performance in both intra-domain and inter-domain recognition tasks. Consequently, these values are designated as the default hyper-parameters in this article.

\begin{figure}[t]
\centerline{\includegraphics[width=80mm]{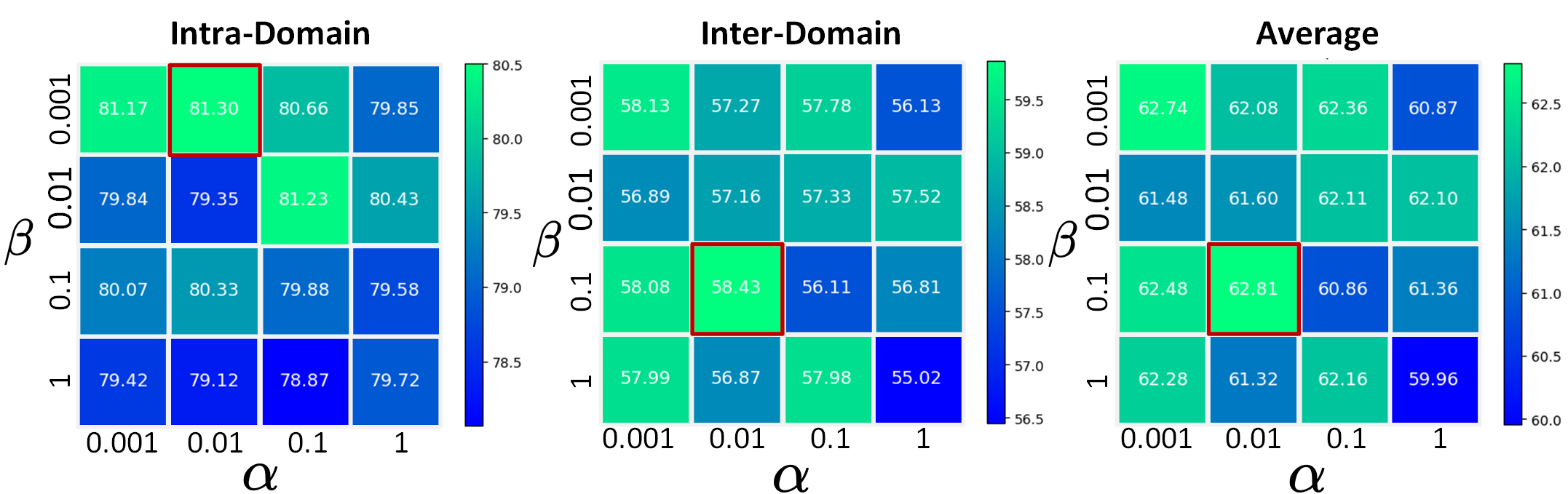}}
\caption[width=80mm]{
The impact of weight hyper-parameters on model performance. The red box represents the optimal performance. 
}
\label{fig2}
\end{figure}

\begin{table}
\caption{Computational complexity comparison results}
\label{table}
\setlength{\tabcolsep}{3pt}
\centering
\begin{tabular}{p{70mm}}
${\includegraphics[width=70mm]{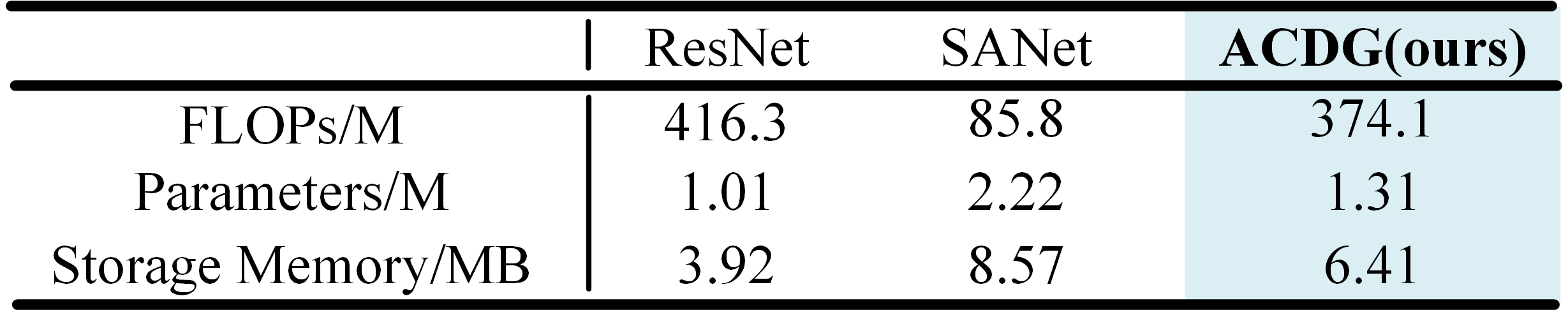}}$
\end{tabular}
\label{table_button}
\end{table}
\subsubsection{Analysis of computational complexity}

Apart from assessing the model's recognition performance, its computational complexity serves as a crucial metric. Consequently, this section delves into reporting the FLOPs (Floating Point Operations), parameter count, and storage memory of the model. These metrics are evaluated from the perspectives of computational complexity, optimization difficulty, and deployment ease, providing a comprehensive evaluation of the model.

The results, as detailed in Table 6, indicate that the proposed ACDG exhibits FLOPs, parameter count, and storage memory of 374.1M, 1.31M, and 6.41MB, respectively. In comparison with existing models such as ResNet and SANet, our method demonstrates no significant increase in computational complexity. However, as previously expounded, our method yields significant improvements in both inter-domain and intra-domain recognition performance. Additionally, it possesses a distinctive capability for spectral denoising, a feature not present in ResNet, SANet, and other existing methods.

\section{Conclusion}

In this paper, we aim to tackle the challenge of inferior performance observed in existing methods when identifying bacterial Raman spectra across different measurement conditions. This issue arises due to the variability in noise levels present in spectra under various measurement conditions. To this end, we introduce a generic ACDG framework designed for the joint denoising and cross-domain identification of bacterial Raman spectra. This method can accomplish spectrum denoising without necessitating any noise-free data. Moreover, it can be extended to address the task of identifying target domain spectra under unobserved measurement conditions, utilizing solely the spectrum from the source domain of a singular measurement condition. We compared it with existing state-of-the-art methods and confirmed its effectiveness. 
Moreover, as a versatile framework, it holds significant potential for application in other biological spectra.

\section{Acknowledgments}
This study was supported in part by the Ministry of Industry and Information Technology of the People's Republic of China. (Grant No. 2023ZY01028).

\printcredits

\bibliographystyle{cas-model2-names}
\bibliography{cas-refs}

\end{document}